\documentclass{IEEEtran}
\usepackage[swedish,english]{babel}
\usepackage{microtype}
\usepackage{graphicx,url,amssymb,steinmetz,bm}
 \usepackage[cmex10]{amsmath}
\usepackage{mathtools}
\usepackage[printonlyused]{acronym}
\acrodef{MIMO}[MIMO]{multiple-input multiple-output}
\acrodef{SISO}[SISO]{single-input single-output}
\acrodef{COI}[COI]{center of inertia}
\acrodef{RHP}[RHP]{right half-plane}
\acrodef{PSSs}[PSSs]{power system stabilizers}
\acrodef{LHP}[LHP]{left half-plane}
\acrodef{LTI}[LTI]{linear time-invariant}
\acrodef{FCR}[FCR]{frequency containment reserves}
\acrodef{FCR-D}[FCR-D]{FCR for disturbance situations}
\acrodef{FFR}[FFR]{fast frequency reserves}
\acrodef{FRR}[FRR]{frequency restoration reserves}
\acrodef{N5}[N5]{Nordic 5-machine}
\acrodef{NMP}[NMP]{nonminimum phase}
\acrodef{DVPP}[DVPP]{dynamic virtual power plant}
\acrodef{MP}[MP]{minimum phase}
\acrodef{MPP}[MPP]{maximum power point}
\acrodef{NREL}[NREL]{National Renewable Energy Laboratory}

\usepackage[hidelinks]{hyperref} 
\usepackage[nameinlink,capitalise]{cleveref}

\crefname{equation}{}{}
\crefname{theorem}{Theorem}{Theorems}
\crefname{prop}{Proposition}{Propositions}
\crefname{cor}{Corollary}{Corollaries}
\crefname{claim}{Claim}{Claims}
\crefname{ex}{Example}{Examples}
\crefname{lemma}{Lemma}{Lemmas}
\usepackage{cite}

\usepackage{bbold}
\usepackage[caption=false,font=footnotesize]{subfig}
\usepackage{siunitx}
\DeclareSIUnit\perunit{p.u.}
\DeclareSIUnit\voltampere{VA}
\DeclareSIUnit\wattsecond{Ws}

\usepackage{accents}

\usepackage{enumitem}

\usepackage[usenames, dvipsnames]{color}

\newtheorem{theorem}{Theorem}

\newtheorem{cor}{Corollary}[theorem]
\newtheorem{lemma}{Lemma}
\newtheorem{rem}{Remark}

\newcommand{\T}{\textit{\textsf{T}}}

\newcommand{\inv}{\ensuremath{{-1}}}
\newcommand{\real}{\ensuremath{\mathbb{R}}}

\newcommand{\complex}{\ensuremath{\mathbb{C}}}
\DeclareMathOperator{\Real}{\ensuremath{\mathrm{Re}}}
\DeclareMathOperator{\Imag}{\ensuremath{\mathrm{Im}}}
\DeclareMathOperator{\diagf}{\ensuremath{\mathrm{diag}}}

\newcommand{\eigmin}{\ensuremath{r}}
\newcommand{\jimag}{  \ensuremath{j} }
\newcommand{\jomega}{  \ensuremath{\jimag \omega} }

\newcommand{\lossyNet}{\ensuremath{\mathcal{L}}}
\newcommand{\transMatrix}{\ensuremath{{W}}}
\DeclareMathOperator{\field}{\ensuremath{\mathrm{F}}}

\newcommand{\Tc}{\textit{\textsf{H}}}

\newcommand{\omegam}{\ensuremath{\bm \omega}}
\newcommand{\deltam}{\ensuremath{\bm \delta}}

\newcommand{\spect}{\ensuremath{{\lambda}}}

\newcommand{\Npoles}{\ensuremath{N}}

\usepackage{tikz}

\usepackage{matlab-prettifier}
\usepackage{scalerel}
\usepackage{tikz}
\usetikzlibrary{svg.path}

\definecolor{orcidlogocol}{HTML}{A6CE39}
\tikzset{
	orcidlogo/.pic={
		\fill[orcidlogocol] svg{M256,128c0,70.7-57.3,128-128,128C57.3,256,0,198.7,0,128C0,57.3,57.3,0,128,0C198.7,0,256,57.3,256,128z};
		\fill[white] svg{M86.3,186.2H70.9V79.1h15.4v48.4V186.2z}
		svg{M108.9,79.1h41.6c39.6,0,57,28.3,57,53.6c0,27.5-21.5,53.6-56.8,53.6h-41.8V79.1z M124.3,172.4h24.5c34.9,0,42.9-26.5,42.9-39.7c0-21.5-13.7-39.7-43.7-39.7h-23.7V172.4z}
		svg{M88.7,56.8c0,5.5-4.5,10.1-10.1,10.1c-5.6,0-10.1-4.6-10.1-10.1c0-5.6,4.5-10.1,10.1-10.1C84.2,46.7,88.7,51.3,88.7,56.8z};
	}
}

\newcommand\orcidicon[1]{\href{https://orcid.org/#1}{\mbox{\scalerel*{
				\begin{tikzpicture}[yscale=-1,transform shape]
				\pic{orcidlogo};
				\end{tikzpicture}
			}{|}}}}

\IEEEoverridecommandlockouts

\begin{document}

\title{A Scalable Nyquist Stability Criterion with Application to Power System Small-Signal Stability}

\author{
\IEEEauthorblockN{Joakim Bj\"ork\IEEEauthorrefmark{1} and Karl Henrik Johansson\IEEEauthorrefmark{2}\IEEEauthorrefmark{3}}
\IEEEauthorblockA{\IEEEauthorrefmark{1} Svenska kraftn\"at,
	Sundbyberg, Sweden}
\IEEEauthorblockA{\IEEEauthorrefmark{2} EECS, KTH Royal Institute of Technology, 
	Stockholm, Sweden}
\IEEEauthorblockA{\IEEEauthorrefmark{3} Digital Futures, 
	Stockholm, Sweden}
\thanks{This work was supported by the KTH Ph.D.\ program in the digitalization of electric power engineering and in part by the Knut and Alice Wallenberg Foundation, the Swedish Research Council, and the Swedish Foundation for Strategic Research.}
\thanks{Joakim Bj\"ork was affiliated with KTH when this work was completed.}
}

\maketitle

\begin{abstract}
    A decentralized stability criterion is derived for a power system with heterogeneous subsystems. A condition for frequency stability and stability of interarea modes is derived using the generalized Nyquist criterion. The resulting scalable Nyquist stability criterion requires only locally available information and gives a priori stability guarantees for connecting new subsystems to an arbitrarily large network. The method can be applied to a general set of agents. For instance, agents with time-delays, nonminimum phase actuators or even unstable dynamics. The scalable Nyquist criterion makes no distinction between nodes with or without synchronous inertia, making it easy to include converter-interfaced renewable energy in the analysis. The method is validated on a detailed nonlinear power system model with frequency droop provided by hydro governors assisted by wind power.
\end{abstract}


\section{Introduction}

With an increasing share of renewable and small-scale generation connecting to the grid, the number of possible power system configurations increases drastically.
Methods addressing global stability has to be scalable and computationally efficient, since the computational effort grows with the system size.
Stability assessment based on centralized computation does not scale well~\cite{patesRobustScalefreeSynthesis2019}, nor does it preserve the privacy of subsystems~\cite{barosDistributedConditionsSmallSignal2021,monshizadehSecantPopovlikeConditions2019,yangDistributedStabilityConditions2020}. Centralized methods are therefore becoming less favorable with the increase in small distributed generation.  A solution to this problem is to instead use decentralized stability conditions. In general, however, decentralized methods come at the cost of conservatism. A careful formulation of the stability criterion is therefore needed to exploit the full potential of the connected devices.

If we do not have any information about the network, we need to make some conservative assumptions to guarantee stability. One solution is to design controllers that ensure passivity of the interconnected system~\cite{yangDistributedStabilityConditions2020,paganiniGlobalAnalysisSynchronization2020}. In general, this is not possible, however, since we may have time delays or zero dynamics that make passivity unachievable. 
In~\cite{faxInformationFlowCooperative2004}, a Nyquist-like criterion is derived for checking the stability of a network of homogeneous \ac{SISO} agents, connected over a static network. In~\cite{gattamiFrequencyDomainCondition2004}, these results are generalized to include networks of homogeneous \ac{MIMO} agents interconnected over a dynamic  network.
Consensus protocols for networks with directed information flow and switching topology have also received attention in the study of self-organizing networked systems~\cite{olfati-saberConsensusCooperationNetworked2007,bulloLecturesNetworkSystems2020}.
For power system applications, however, we are concerned with fixed networks. 
In~\cite{patesRobustScalefreeSynthesis2019}, a robust scale-free synthesis method is developed, guaranteeing stability by identifying a separating hyperplane in the Nyquist diagram.
The method provides a priori stability guarantees for connecting new devices to the grid. 
In this paper, we will present a generalization of the results in~\cite{patesRobustScalefreeSynthesis2019} and~\cite{faxInformationFlowCooperative2004} using the generalized Nyquist criterion in combination with the field of values. 

The main contribution of this work is a scalable Nyquist stability criterion allowing for a network of heterogeneous agents coupled over a connected (possibly lossy) network. 
When introducing this novel method, we make the common assumption  that the network is first-order~\cite{barosDistributedConditionsSmallSignal2021,monshizadehSecantPopovlikeConditions2019,patesRobustScalefreeSynthesis2019,paganiniGlobalAnalysisSynchronization2020,kundurPowerSystemStability1994}. That is, we allow for an arbitrarily large network, but the dynamics at each node are \ac{SISO}. In the power system application, this means that we only model the phase angle and active power dynamics, neglecting the voltage and reactive power dynamics.
By directly applying the generalized Nyquist criterion, we allow for a general set of \ac{LTI} agents. In the paper, we distinguish between exponential stability and asymptotic synchronization on the average network mode. For a system to be exponentially stable, we require asymptotic synchronization, but also that the average mode is stable~\cite{bulloLecturesNetworkSystems2020}. In power systems, we are only concerned about the average frequency mode, i.e., we only require the derivative of the average mode to be
stable. 
The result is a general analysis framework for assessing power system stability, applicable both to conventional thermal and hydro units, as well as converter-interfaced generation such as wind and solar.
The results are validated in detailed nonlinear power system simulations in a 5-machine test system modeled after the Nordic grid.
Local stability criteria are derived for heterogeneous networks with time-delayed actuators and \ac{NMP} hydro units and wind turbines participating in \ac{FCR} and \ac{FFR}, respectively. The benefit of the proposed method is that it allows for a very general set of agents and dynamics. The criterion allows us to formulate a stability criterion for agents with time-delayed actuators in combination with uncontrolled agents, something that is not possible using methods based solemnly on passivity or a separating hyperplane in the Nyquist diagram. The method also allows for nodes with no inertia, such as converter-interfaced renewable energy. The proposed scalable Nyquist stability criterion can also be applied in situations where we have unstable agent dynamics, something that can easily occur if we have realistic actuator dynamics.

The remainder of this paper is organized as follows. 
\cref{sec:hetero:prel}, introduces the generalized Nyquist criterion and field of values.  \cref{sec:power_syst_model}, introduces the network model and \cref{sec:class_net_stab}, presents the classification of network stability. \cref{sec:scalable_Nyquist} presents the main result: a criterion that guarantee stability of interarea modes using only local information. In \cref{sec:power_system_app}  the results are validated in detailed nonlinear power system simulations. \cref{sec:hetero:concl} concludes the paper.

%%%%%%%%%%%%%%%%%%%%%%%%%%%%%%%%%%%%%%%%%%%%%%%%%%%%%%%%%%%
\section{Preliminaries}
\label{sec:hetero:prel}
We review some results for \ac{MIMO} \ac{LTI} systems ~\cite{maciejowskiMultivariableFeedbackDesign1989,postlethwaiteComplexVariableApproach1979,skogestadMultivariableFeedbackControl2007,zhouRobustOptimalControl1996,brownComplexVariablesApplications2004}.	
Let $L(s)$, $s\in\complex$, denote a square, proper, and rational transfer matrix with no internal \ac{RHP} pole-zero cancellations. Assume that the feedback system with return ratio $-L(s)$ is well posed. Let $\det\big(I+L(s)\big) = \frac{\phi_{cl}(s)}{\phi_{ol}(s)}$, where $\phi_{ol}(s)$ and $\phi_{cl}(s)$ are the open- and closed-loop characteristic polynomials, respectively. The closed-loop system is stable if and only if $\phi_{cl}(s)$ have no roots in the \ac{RHP}.
Define the Nyquist $\mathcal{D}$-contour as a contour in the complex plane that includes the entire $\jomega$-axis and an infinite semi-circle into the \ac{RHP}, making small indentations into the \ac{RHP} to avoid any open-loop poles of $L(\jomega)$ (roots of $\phi_{ol}(s)$) directly on the $\jomega$-axis.

\begin{lemma}[%Generalized Nyquist Criterion~
Generalized Nyquist Criterion~\cite{maciejowskiMultivariableFeedbackDesign1989}]%
	\label{lemma:gen_MIMO}%	
	If $L(s)$ has $\Npoles$ unstable (Smith-McMillan) poles, then the closed-loop system with return ratio $-L(s)$ is stable if and only if the eigenloci of $L(s)$, taken together, encircle the point $-1$ $\Npoles$ times anticlockwise, as $s$ goes \mbox{clockwise around the Nyquist $\mathcal{D}$-contour.}	
\end{lemma}

The spectrum of a complex matrix $A\in\complex^{n\times n}$ is the set of eigenvalues $\spect(A)=\{\spect_1(A),\ldots,\spect_n(A)\}$. 
The spectrum lies inside the field of values
$
\field(A) := \left\{v^\Tc A v : v \in \complex^n,v^\Tc v =1 \right\}
$ see~\cite{hornTopicsMatrixAnalysis1991}.
Let $B\in\real^{n\times n}$ be a positive semi-definite matrix with $0\leq \spect_1(B)\leq\cdots\leq \spect_n(B)$.
Then the $ij$\textsuperscript{th} element of the product
$\spect_1(B)(A_{ij}) \leq (AB)_{ij} \leq \spect_n(B)(A_{ij})$.
Consequently,
\begin{multline}
\label{eq:field_product}
\spect\big(A B\big) \subset \field \big(A B\big) =  \alpha\field \big(A\big) 
\\
= \left\{v^\Tc A v : v \in \complex^n,v^\Tc v = \alpha, \alpha \in [\spect_1(B),\spect_n(B)]\right\}.
\end{multline}

\section{Power System Model}
\label{sec:power_syst_model}
\begin{figure}[t!]
	\centering
	\includegraphics[scale=1.2]{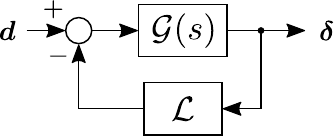}
	\caption{Network of $n$ agents coupled through the network $\mathcal{L}$.}
	\label{fig:blocksimplenetwork_hetero}
\end{figure}

We consider the phase angle dynamics in a power system with $n$ buses. The dynamics at bus $i\in\{1,\ldots,n\}$ can be described by the swing equation
\begin{equation}
\label{eq:swing}
    \begin{aligned}
    s \delta_i(s) &= \omega_i(s)
    \\
    s M_i \omega_i(s) & = - R_i(s)\delta_i(s) -F_i(s) \omega_i(s) + d_i(s)
    \end{aligned}
\end{equation}
where $\delta_i(s)$ represent the voltage phase angle, $\omega_i(s)$ the frequency, and $d_i(s)$ represent some external power input at bus $i$. 
The \ac{SISO} transfer functions $R_i(s)$ and $F_i(s)$ represent local phase angle and frequency dependent actuators, respectively, whereas the constant $M_i\geq 0$ represents the inertia. If $M_i>0$, then the agent represents a synchronous machine.
The transfer function from $d_i$ to $\delta_i$ can be written as 
\begin{equation}
\label{eq:def_agent}
g_i(s) := \frac{1}{s^2M_i + s F_i(s) + R_i(s)}
\end{equation}
for instance, representing the dynamics of a synchronous machine with or without governor, a load, or a power electronics device.
To improve readability,  we do not write out signal dependency on $s\in \complex$, e.g., we let $\delta_i(s) = \delta_i$, $\omega_i(s) = \omega_i$, and $d_i(s) = d_i$ in the remainder of the paper.
The closed-loop network, shown in \cref{fig:blocksimplenetwork_hetero}, can be described by 
\begin{equation}
	\label{eq:hetero_feedback1}
		\deltam = \mathcal{G}(s)(\bm{d}-\mathcal{L} \deltam)
\end{equation}
where outputs $\deltam = [\delta_1,\ldots,\delta_n]^\T$  and $\bm{d} = [d_1,\ldots,d_n]^\T$. 
The agent dynamics ${\mathcal{G}}(s) = \diagf\big({g}_1(s),\ldots,{g}_n(s)\big)$ are 
coupled through a first-order network described by the matrix~$\mathcal{L}$.

For the analysis, we make the standard assumptions that the bus voltage magnitudes are constant for the time frame of interest, the transmission is lossless, and that reactive power does not affect the voltage phase angles~\cite{kundurPowerSystemStability1994}.
Then, $\mathcal{L}$ is a Laplacian matrix with elements
\begin{equation}
\label{eq:pates_edges}
\mathcal{L}_{ij} = \frac{\partial}{\partial \delta_j} \sum\nolimits_{l=1}^{n} V_{i}^*V_{l}^*b_{il}\sin(\delta_i-\delta_l)\Big |_{\deltam=\deltam^*},
\end{equation}
where $\bm{\delta}^*$ and $V_i^*$ represent the phase angles and voltage magnitudes, respectively, at the linearization point, and $b_{il}\geq 0$ is the susceptance of the transmission line connecting buses $i$ and $l$. If $b_{il}=0$, then buses are not directly connected~\cite{patesRobustScalefreeSynthesis2019}.

Equivalently, \eqref{eq:hetero_feedback1} can be written as
\begin{equation}
\label{eq:hetero_feedback2}
s E \begin{bmatrix}
\deltam \\
\omegam
\end{bmatrix} = A(s) \begin{bmatrix}
\deltam \\
\omegam
\end{bmatrix} + B \bm d, \quad E = \diagf(I,\mathcal{M}),
\end{equation}
where
\begin{equation}
\label{eq:hetero_feedback3}
A(s) = \begin{bmatrix}
0 &  I \\
-\mathcal{L}-\mathcal{R}(s) & -\mathcal{F}(s)
\end{bmatrix} 
, \quad
B = \begin{bmatrix}
0 \\
I
\end{bmatrix},
\end{equation}
the transfer matrices $\mathcal{R}(s) = \diagf(R_1(s), \ldots, R_n(s))$ and $\mathcal{F}(s) = \diagf(F_1(s), \ldots, F_n(s))$, and the constant matrix $\mathcal{M} = \diagf(M_1, \ldots, M_n)$. The vector $\bm \omega= [\omega_1,\ldots,\omega_n]^\T= s \deltam$ represent the voltage frequency at each node.
We assume that there are no algebraic network nodes, i.e., there are no node $i$ such that $M_i=F_i(s)=R_i(s)=0$. This is not a restriction since we can always formulate a reduced network model without algebraic nodes by taking the Schur complement of $\mathcal{L}$ with respect to the algebraic nodes. This reduction of an electrical network is known as Kron reduction~\cite{dorflerKronReductionGraphs2013}. 

Let the eigenvalue decomposition of $\mathcal{L}$ be
\begin{equation}
\label{eq:eigendecomp_hetero}
\diagf \! \big(\spect_1(\mathcal{L}),\ldots,\spect_n(\mathcal{L})\big) =\diagf(\lambda_1,\ldots,\lambda_n) = \mathcal{V}^\T \mathcal{L} \mathcal{V}
\end{equation}
where
$\mathcal{V}\in\real^{n\times n}$ is a unitary matrix of eigenvectors $\mathcal{V} = \begin{bmatrix}
v_1,\ldots,v_n
\end{bmatrix}$ so that $\mathcal{V}^\T \mathcal{V} = I$.
Let the eigenvalues be arranged in ascending order so that $0 = \lambda_1 < \lambda_2 \leq \cdots \leq \lambda_n$.
Since $\mathcal{L}$ is a Laplacian matrix, $\lambda_1 = 0$ and the corresponding eigenvector $v_1 = \bm{1}/\sqrt{n}$, where  $\bm{1}$ is a vector of ones. The mode $\varLambda_1 = v_1^\T \deltam$ describes the average phase angle, whereas  $\varOmega_1 = v_1^\T \omegam$ describes the average frequency of the network.

\section{Classification of Network Stability}
\label{sec:class_net_stab}
In this section we present the classification of network stability used in this paper.
A common way to characterize the stability of a power system is by diagonalizing the system equations~\cite{kundurPowerSystemStability1994}. System stability is then expressed in terms of the stability of network modes, for example, the average mode plus the interarea modes. 

Consider the coordinate change to modal states
\begin{equation}
\label{eq:coord_trans_mode}
\begin{bmatrix}
\varLambda_1,  \varOmega_1, \ldots,  \varLambda_n,  \varOmega_n
\end{bmatrix}^\T := \transMatrix^\T \begin{bmatrix}
\deltam \\ \omegam
\end{bmatrix} 
\end{equation}
using the transformation matrix
\begin{equation}
\transMatrix := \begin{bmatrix}
v_1 & 0 & \cdots & v_n & 0\\
0 & v_1 & \cdots & 0 & v_n\\
\end{bmatrix}\in \real^{2n \times 2n}.
\end{equation}
% made up of the eigenvectors $v_i$, $i\in\{1,\ldots,n\}$ of $\mathcal{L}$. 
Since $\transMatrix^\T = \transMatrix^\inv$, the coordinate transform \cref{eq:coord_trans_mode} applied to \cref{eq:hetero_feedback2},
\begin{equation}
\label{eq:sim_transform}
s \transMatrix^\T E \transMatrix \transMatrix^\T \! \begin{bmatrix}
\deltam \\ \omegam
\end{bmatrix} \! = \transMatrix^\T A(s) \transMatrix \transMatrix^\T \! \begin{bmatrix}
\deltam \\ \omegam
\end{bmatrix} \! + \transMatrix^\T B \mathcal{V} \mathcal{V}^\T \bm d,
\end{equation} 
is a similarity transformation. If the network is homogeneous, then \eqref{eq:sim_transform} is a block-diagonal realization of \cref{eq:hetero_feedback2}, with the $2\times 2$ blocks
\begin{equation}
	\setlength\arraycolsep{3pt}
s \! \begin{bmatrix}
\varLambda_i \\  M_{\lambda_i}\varOmega_i 
\end{bmatrix} 
\!=\! 
\begin{bmatrix}
0 & 1\\-\lambda_i \! -\! R_{\lambda_i}(s)  & - F_{\lambda_i}(s)
\end{bmatrix} 
\!
\begin{bmatrix}
\varLambda_i \\ \varOmega_i 
\end{bmatrix} 
+
\begin{bmatrix}
0 \\ 1 
\end{bmatrix} \! v_i^\T \bm{d},
\end{equation}
characterizing the dynamics of network mode~$i$, where $M_{\lambda_i} = v_i^\T \mathcal{M} v_i$, $F_{\lambda_i}(s) = v_i^\T \mathcal{F}(s) v_i$, and $R_{\lambda_i}(s) = v_i^\T \mathcal{R}(s) v_i$.

The transfer function from $v_i^\T \bm d$ to $\varLambda_i$ is 
\begin{equation}
\label{eq:dlambda_i-to-deltalambda_i}
\frac{1}{s^2 M_{\lambda_i} + sF_{\lambda_i}(s) + R_{\lambda_i}(s) + \lambda_i} 
= \frac{h_i(s)}{1 + {\lambda_i} h_i(s)},  
\end{equation} 
where
\begin{equation}
h_i(s) := \frac{1}{s^2M_{\lambda_i}  + s F_{\lambda_i}(s) + R_{\lambda_i}(s)}.
\end{equation}
Since the similarity transform preserves stability, the network~\eqref{eq:hetero_feedback1} is stable if  \eqref{eq:dlambda_i-to-deltalambda_i} is stable for all $i\in\{1,\ldots,n\}$. 
We can apply the Nyquist criterion on the \ac{SISO} return ratios~$-{\lambda_i} h_i(s)$ to see how each agent affects the network modes. For a heterogeneous network, however, stability of \eqref{eq:dlambda_i-to-deltalambda_i}  only approximately relates to the stability of interarea modes, $i\in\{2,\ldots,n\}$.
In \cref{sec:scalable_Nyquist}, we derive a more general criterion that can be applied also to a heterogeneous network, at the cost of being more conservative.

Note that it is possible to characterize the stability of the average mode, $i=1$, also for the heterogeneous case, since we know that $v_1 = \bm{1}/\sqrt{n}$.
The transfer function  of the average frequency mode is
\begin{equation}
	\label{eq:reduced_avg_freq}
	\omega_\mathrm{avg} :=
	\varOmega_1/\sqrt{n}=  \frac{s}{s^2 M+ s F(s) +  R(s)} \sum\nolimits_{i=1}^n d_i,
\end{equation}
where $M = \sum\nolimits_{i=1}^n M_i$, $F(s) = \sum_{i=1}^n F_i(s)$, and $R(s) = \sum_{i=1}^n R_i(s)$.
Note that we can incorporate network losses and any phase angle dependent actuators in $\mathcal{R}(s)$ into $\mathcal{F}(s)$ by substituting $\mathcal{F}(s)$ with $\mathcal{F}(s)+\frac{1}{s}\mathcal{R}(s)$ in \eqref{eq:hetero_feedback3}.
Thus, with the first-order network model, the average frequency disturbance response is given by
\begin{equation}
\label{eq:COI_freq_dist_response_first_order}
\omega_\mathrm{avg}  = \frac{1}{sM + F(s)} \sum\nolimits_{i=1}^n d_i.
\end{equation}

We classify the stability of the closed-loop network \eqref{eq:hetero_feedback1} using the modal states \eqref{eq:coord_trans_mode}. 
\begin{itemize}
	\item If modes $i\in\{1,\ldots,n\}$ are stable, then the system is exponentially stable.
	\item If modes $i\in\{2,\ldots,n\}$ are stable, then  the system achieves asymptotic synchronization on the average mode (the system has stable interarea modes). Exponential stability therefore implies asymptotic synchronization.
\end{itemize}
We say that the network \eqref{eq:hetero_feedback1} has stable frequency dynamics if the average frequency \eqref{eq:COI_freq_dist_response_first_order} and interarea modes are stable.

\begin{rem}
	\label{rem_COI_differ}
If we have homogeneous or proportional agents, then $\omega_\mathrm{avg}$ is equal to the \ac{COI} frequency 
\begin{equation}
	\label{eq:COI_standard}
	\omega_\mathrm{COI} := {\sum\nolimits_{i=1}^n \omega_i M_i}\big/{M}.
	%\frac{\sum_{i=1}^n M_i\omega_i}{M}.
\end{equation}
For heterogeneous agents, the exact representation of the \ac{COI} mode cannot easily be obtained~\cite{paganiniGlobalAnalysisSynchronization2020}. The problem is that the \ac{COI} mode contains information about the higher-order network modes, i.e., the interarea modes. This makes the transient response of \eqref{eq:COI_standard} different from \eqref{eq:COI_freq_dist_response_first_order}. However, if the system achieves asymptotic synchronization on the average mode then $\omega_\mathrm{COI}$ and $\omega_\mathrm{avg}$ converge.
\end{rem}

\section{Scalable Nyquist Stability Criterion}
\label{sec:scalable_Nyquist}
\begin{figure}[t]
	\centering
	\includegraphics[width=\linewidth]{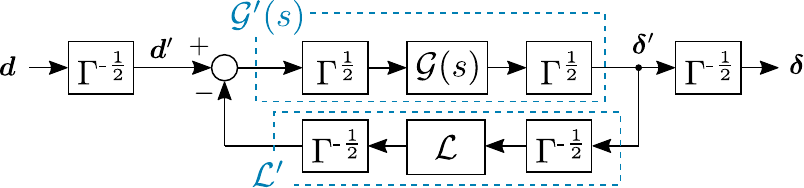}
	\caption{Network normalized using \cref{eq:weighted_gamma}.}
	\label{fig:blocknetworknormal}
\end{figure}

Let
\begin{equation}
\label{eq:weighted_gamma}
\mathcal{G}'(s) := \Gamma^{\frac{1}{2}} \mathcal{G}(s) \Gamma^{\frac{1}{2}}, \quad \text{and} \quad \mathcal{L}' := \Gamma^{\text{-}\frac{1}{2}} \mathcal{L} \, \Gamma^{\text{-}\frac{1}{2}},
\end{equation}
where $\Gamma$ is the diagonal matrix
\begin{equation}
	\label{eq:incidence parameters}
\Gamma :=\diagf(\gamma_1,\ldots,\gamma_n) = 2\cdot\diagf(\mathcal{L}_{11},\ldots,\mathcal{L}_{nn})
\end{equation}
with $\mathcal{L}_{ii}$ being the diagonal entries of $\mathcal{L}$.
The eigenvalues of $\mathcal{L}'$ are then $0=\mu_1 < \mu_2\leq \cdots \leq \mu_n \leq 1$, where $\mu_2$ is the algebraic connectivity of $\mathcal{L}'$~\cite{bulloLecturesNetworkSystems2020}.
Stability of the interconnection of $\mathcal{G}(s)$ over $\mathcal{L}$ is equivalent to stability of the normalized interconnection of $\mathcal{G}'(s)$ over $\mathcal{L}'$, as shown in \cref{fig:blocknetworknormal}. 
The network \eqref{eq:hetero_feedback1} is exponentially stable if the transfer functions
	\begin{equation}
	\label{eq:thm_lossy_1}
	\begin{aligned}
	\mathcal{G}'(s) \mathcal{L}'\big(I+\mathcal{G}'(s) \mathcal{L}'\big)^\inv, \quad \big(I+\mathcal{G}'(s) \mathcal{L}'\big)^\inv
	\\
	\mathcal{G}'(s) \big(I+\mathcal{G}'(s) \mathcal{L}'\big)^\inv, \quad \mathcal{L}' \big(I+\mathcal{G}'(s) \mathcal{L}'\big)^\inv
	\end{aligned}
	\end{equation}
	are all stable. 

\subsection{Exponential Stability}
Consider first the special case where we assume that the network is lossy\footnote{
	Since we ignore the voltage and reactive power dynamics, there is no physical meaning of network losses. 
	However, they are conceptually the same as frequency-dependent controlled actuators with integral action. 
}, e.g., substitute the matrix $\mathcal{L}'$ with $\mathcal{L}' + \epsilon I$, $\epsilon>0$. 
	Since the new $\lossyNet'$ is constant and has full rank, it is sufficient to check one of the four transfer functions in \eqref{eq:thm_lossy_1}.	
	Factorize $\lossyNet' = {\mathcal{W}} {Y} {\mathcal{W}}^\T$, where the diagonal matrix $Y>0$, and ${\mathcal{W}}^\T {\mathcal{W}} = I$. 
	We have that
	\begin{align}
	\lossyNet'(I+\mathcal{G}'(s) \lossyNet')^\inv &=
	{\mathcal{W}} {Y} {\mathcal{W}}^\T(I+\mathcal{G}'(s) {\mathcal{W}} {Y} {\mathcal{W}}^\T)^\inv \nonumber\\
	&= {\mathcal{W}} {Y} (I+{\mathcal{W}}^\T \mathcal{G}'(s) {\mathcal{W}} {Y} )^\inv{\mathcal{W}}^\T.	\label{eq:proof:decompose1}
	\end{align}
	Clearly, \eqref{eq:proof:decompose1} is stable if the $n\times n$ sensitivity function
	\begin{equation}
	\label{eq:proof:decompose2}
	\big(I + P(s)\big)^\inv  : =\big(I + {\mathcal{W}}^\T \mathcal{G}'(s){\mathcal{W}} {Y}\big)^\inv 
	\end{equation}
	is stable. Since the feedback system \cref{eq:proof:decompose2} with return ratio $-P(s)$ is well posed, we can assess stability using \cref{lemma:gen_MIMO}. Let $\Npoles$ be the number of unstable poles in $\mathcal{G}'(s)$.
	The closed-loop \eqref{eq:proof:decompose2} is then stable if and only if the image of
\begin{align}
	\label{eq:thm_lossy_3}
	\det\big(I+P(s)\big) &= \prod\nolimits_{i=1}^{n} \spect_i\big(I+P(s)\big) \nonumber \\&= \prod\nolimits_{i=1}^{n} \Big(1 + \spect_i\big(P(s)\big)\Big)
\end{align}	
	makes $\Npoles$ anticlockwise encirclements of the origin as $s$ goes clockwise around the Nyquist $\mathcal{D}$-contour. Note that the image of $1 + \spect_i\big(P(s)\big)$ encircles the origin if $\spect_i\big(P(s)\big)$ encircles the point $-1$; and that the argument of a product is the same as the sum of the arguments. The closed-loop system \eqref{eq:proof:decompose2} is therefore  stable if and only if the eigenloci of $P(s)$, taken together, encircle the point $-1$ $\Npoles$ times.

\subsection{Asymptotic Synchronization}
\begin{figure}
		\centering
		\includegraphics[width=\linewidth]{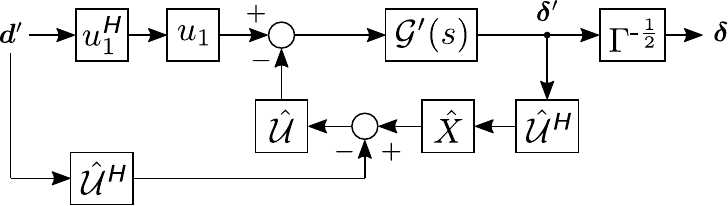}
		\caption{Normalized network with separated average and interarea modes.}
		\label{fig:blockmodes}
	\end{figure}

Consider now the power system example introduced in \cref{sec:power_syst_model}. Here, the  closed-loop system has a lossless network matrix with $\spect_1(\mathcal{L}') = 0$. The average frequency mode is not controllable over the network since $\mathcal{L}'u_1 = 0$.
Thus, the feedback system is ill-posed and we cannot directly apply \cref{lemma:gen_MIMO} to assess stability of the system. Instead, we first separate the average mode from the interarea modes.
Factorize $\mathcal{L}' = \mathcal{U} X \mathcal{U}^\T = \hat{\mathcal{U}} \hat X \hat{\mathcal{U}}^\T$.
Normalized using \cref{eq:weighted_gamma},
$0 \leq X \leq I \in \real^{n\times n},$ and $ \mathcal{U} = [u_1,\ldots,u_n]\in \real^{n\times n}$, while
$\mu_2 I \leq \hat X \leq I \in \real^{n-1\times n-1}$, and $\hat{\mathcal{U}} = [u_2,\ldots,u_n]$
%The factorization separates the average mode from the interarea modes. 
such~that
\begin{equation}
\mathcal{U} \mathcal{U}^\T = u_1 u_1^\T + \hat{\mathcal{U}} \hat{\mathcal{U}}^\T = I \in \real^{n\times n}.
\end{equation}	
The transfer function from $\bm{d}'$ to $\deltam'$ is therefore equivalent for the closed-loop systems shown in \cref{fig:blocknetworknormal} and \cref{fig:blockmodes}.

Stability of the interarea modes can be assessed using
\begin{equation}
\mathcal{L}' \big(I+\mathcal{G}'(s) \mathcal{L}'\big)^\inv = \hat{\mathcal{U}} \hat{X} (I+\hat{\mathcal{U}}^\T \mathcal{G}'(s) \hat{\mathcal{U}} \hat{X} )^\inv\hat{\mathcal{U}}^\T.
\end{equation}
The interarea modes		
are stable if the $(n-1)\times (n-1)$ sensitivity
\begin{equation}
\label{eq:proof:decompose3}
\big(I + L(s)\big)^\inv  : =\big(I+\hat{\mathcal{U}}^\T \mathcal{G}'(s) \hat{\mathcal{U}} \hat{X}\big)^\inv 
\end{equation}
is stable. Since the feedback system with return ratio $-L(s)$ is well posed, we can assess stability using \cref{lemma:gen_MIMO}.
The sensitivity \eqref{eq:proof:decompose3} is stable if 
\begin{equation}
\label{eq:thm_lossy_4}
\det\big(I+L(s)\big) = \prod\nolimits_{i=1}^{n-1}\Big( 1 + \spect_i\big(L(s)\big) \Big)
\end{equation}	
makes $\Npoles$ anticlockwise encirclements of the origin as $s$ goes clockwise around the Nyquist $\mathcal{D}$-contour. That is, if the eigenloci of $L(s)$, taken together, encircle the point $-1$ $\Npoles$ times. This gives us the following result:

\begin{theorem}[Asymptotic Synchronization Criterion]%
	\label{thm:syncronization_hetero}%
	Assume that $\mathcal{G}'(s)$ has $\Npoles$ unstable poles. Then the closed-loop system with return ratio $-\mathcal{L}' \mathcal{G}'(s)$ achieves asymptotic synchronization on the average mode if and only if the eigenloci $
	\left\{\spect_2\big(\mathcal{L}' \mathcal{G}'(s)\big),\ldots,\spect_n\big(\mathcal{L}' \mathcal{G}'(s)\big)\right\}
	$, taken together,
	encircle the point $-1$ $\Npoles$ times anticlockwise, as $s$ goes clockwise around the Nyquist $\mathcal{D}$-contour.
\end{theorem}

If we assume that $\Npoles = 0$, then we can formulate a conservative stability criterion using \eqref{eq:field_product}.
Note that 
	\begin{equation}
	\spect_i\big(\mathcal{L}' \mathcal{G}'(s)\big) \in \spect\big(\hat{\mathcal{U}}^\T \mathcal{G}'(s) \hat{\mathcal{U}} \hat{X}\big) , \quad \forall i \in\{2,\ldots,n\},
	\end{equation}
	where $\hat{X} = \diagf(\mu_2,\ldots,1)$. Consequently, \eqref{eq:field_product} gives
	\begin{equation}
	\spect\big(\hat{\mathcal{U}}^\T \mathcal{G}'(s) \hat{\mathcal{U}} \hat{X}\big) \subset \alpha \field\big(\hat{\mathcal{U}}^\T \mathcal{G}'(s) \hat{\mathcal{U}}\big) = \alpha \field\big(\mathcal{G}'(s)\big)
	\end{equation}
	where $\alpha \in [\mu_2,1]$, with $\mu_2= \lambda_2(\mathcal{L}')$ being the algebraic connectivity of the network $\mathcal{L}'$.
	Stability of the interarea modes can then be assessed using \cref{eq:proof:decompose3,eq:thm_lossy_4},
	noting that if the field of values does not include or encircle the point $-1$, then the eigenloci cannot encircle $-1$. This gives us the paper's main result:

\begin{cor}[Scalable Nyquist Stability Criterion]%
\label{cor:syncronization_hetero}%
Assuming that $\mathcal{G}'(s)$ has no unstable poles, then asymptotic synchronization on the average mode is guaranteed if 
the field of values
\begin{equation}
\label{eq:field_value_cor}
\alpha \field\big(\mathcal{G}'(s)\big), \quad \alpha \in (0,1]
\end{equation}
does not encircle  $-1$ as $s$ goes around the Nyquist $\mathcal{D}$-contour.
\end{cor}

\subsection{Stability of Interarea Modes}
With realistic governor dynamics, $\mathcal{G}'(s)$ may very well have unstable poles, as we will see in \cref{sec:power_system_app} later on. However, if we are concerned with the stability of interarea modes, then we are only interested in unstable closed-loop poles in the frequency range of the interarea modes. 
Typically, we at least have a good idea about the frequency of the slowest interarea mode. 
Assume that we know that the frequency of the slowest interarea mode is bounded from below by $\eigmin>0$ for all possible operating conditions. Let $\mathcal{D}_\eigmin$ be the modified Nyquist contour with an indentation into the \ac{RHP} with radius $\eigmin$ at the origin. 
Then we can formulate a relaxed version of \cref{cor:syncronization_hetero}:
\begin{cor}[Relaxed Scalable Nyquist Stability Criterion]%
\label{cor:relax_syncronization_hetero}
    Assuming that $\mathcal{G}'(s)$ does not have unstable poles inside the \mbox{$\mathcal{D}_\eigmin$-contour}, then the closed-loop system with return ratio $-\mathcal{L}'\mathcal{G}'(s)$ has stable interarea modes if
the field of values \eqref{eq:field_value_cor}
does not encircle $-1$ as $s$ goes around the \mbox{$\mathcal{D}_\eigmin$-contour.}
\end{cor}

\section{Power System Application}
\label{sec:power_system_app}
In this section we will show how to formulate a decentralized stability criterion in a realistic power system with \ac{NMP} actuators and time-delays. First, we introduce the \ac{N5} test system and the nonlinear models of hydro units and wind turbines, and their linearizations. Then we show how to derive a decentralized stability criterion in a network with only hydro--\ac{FCR} and for a network with both hydro--\ac{FCR} and wind--\ac{FFR}, using \cref{cor:relax_syncronization_hetero}.

\subsection{The Nordic 5-Machine Test System}
\begin{figure}[t]
	\includegraphics[width=1\linewidth]{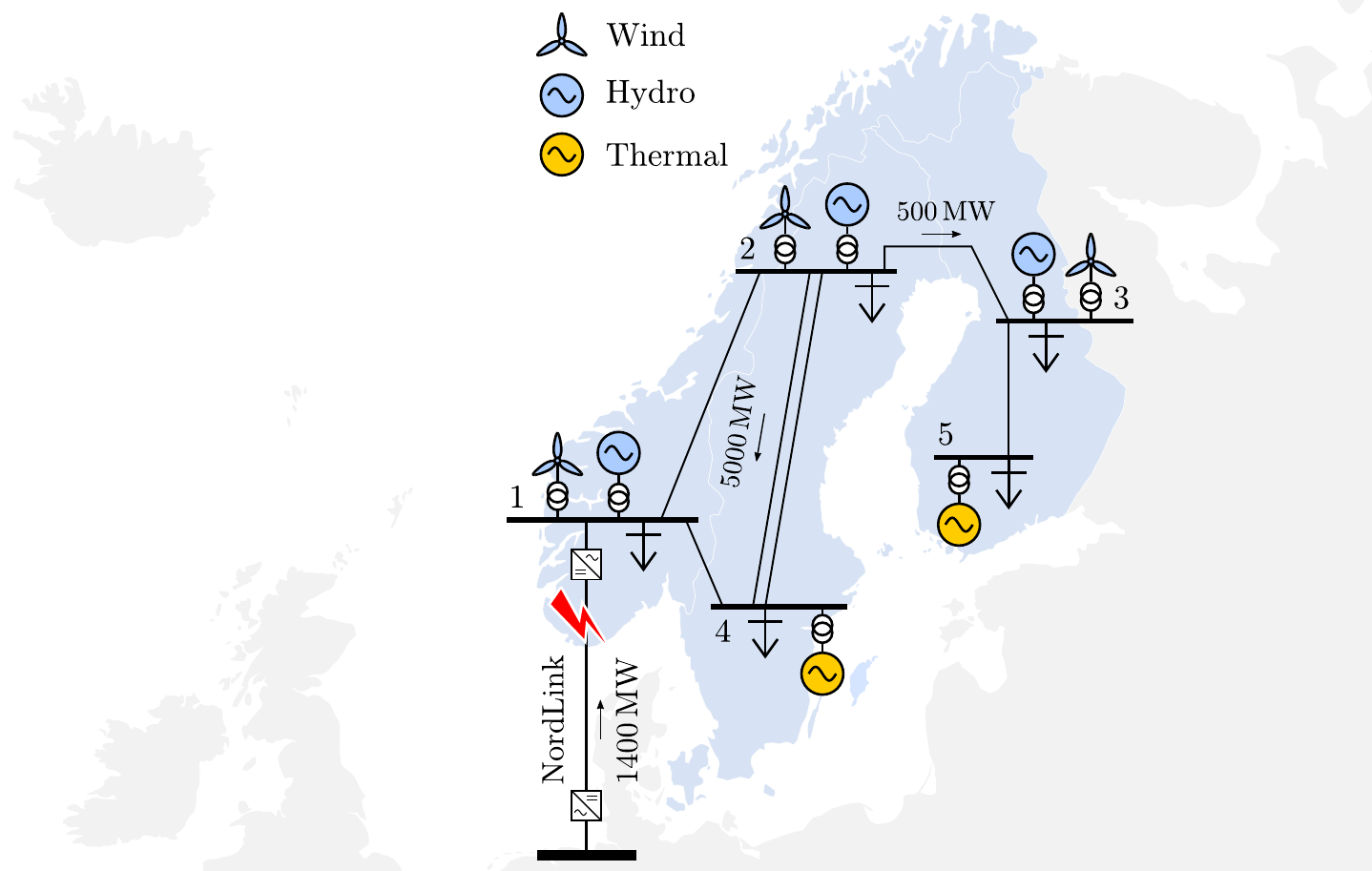}
	\caption{One-line diagram of the N5 test system. The system is implemented in Simulink and is available at \url{https://github.com/joakimbjork/Nordic5}. }
	\label{fig:nordicedit}
\end{figure}

Consider the \ac{N5} test system shown in \cref{fig:nordicedit}. The system was developed in~\cite{bjorkDynamicVirtualPowerunpublished,bjorkVariablespeedWindTurbineunpublished,bjorkFundamentalControlPerformance2021} to study the coordination of slow \ac{FCR} from hydro with \ac{FFR} from wind in a low-inertia power system. The system is phenomenological but has dynamic properties similar to those of the Nordic synchronous grid. 
% The test system is  implemented in Simulink Simscape Electrical. %\cite{hydro-quebecSimscapeElectricalReference2020_manual}.
Loads and machines are lumped up into a single large unit at each bus. The hydro and thermal units are modeled as 16\textsuperscript{th} order salient-pole and round rotor machines, respectively. 

In the Nordic system, the frequency of the slowest interarea mode can be expected to be around \SI{0.4}{\hertz}, depending on the operating condition. In the \ac{N5} test system, the slowest interarea mode (the mode between buses 1, 4 and buses 2, 3, and 5) ranges from \SI{0.37}{\hertz} during high-inertia operating conditions, to \SI{0.49}{\hertz} during low-inertia operating conditions. 

The Nordic system currently applies two types of \ac{FCR}: \ac{FCR} for normal operation, within $\text{50.0}\pm\text{\SI{0.1}{\hertz}}$; and \ac{FCR-D}, activated when the frequency falls below \SI{49.9}{\hertz}. \ac{FCR-D} have a faster response time and is designed to limit the maximum instantaneous frequency deviation to \SI{49.0}{\hertz}, and to stabilize the system at \SI{49.5}{\hertz}~\cite{entso-eNordicSynchronousArea2017}.

The kinetic energy of the system varies greatly over the year, since the amount of synchronous generation connected to the grid depends on the demand \cite{entso-eFastFrequencyReserve2019}.
For this analysis, we consider a low-inertia scenario with $W_\mathrm{kin} = \SI{110}{\giga\wattsecond}$ distributed according to \cref{tab:Nordic5_2}. 
Assume that we have constant power loads with a combined frequency dependency of \SI{400}{\mega\watt\per\hertz} and consider the dimensioning fault to be the loss of a \SI{1400}{\mega\watt} importing dc link. The \ac{FCR-D} requirements are then fulfilled in the average frequency model \eqref{eq:reduced_avg_freq} if the total \ac{FCR} amount to
\begin{equation}
	\label{eq:desiredKfcr}
	F_\mathrm{des}(s) = k \frac{6.5s+1}{(2s+1)(17s+1)},
\end{equation}
where $k=\SI{3100}{\mega\watt\per\hertz}$~\cite{bjorkDynamicVirtualPowerunpublished}. For the analysis, we let the bus dynamics be 
\begin{equation}
	\label{eq:agent_Nordic5}
	g_i(s) = \frac{1}{s} \frac{1}{s M_i+ F_i(s) + D_i}, \quad i\in\{1,\ldots,5\},
\end{equation}
where $M_i\geq 0$ is the inertia\footnote{At nominal frequency, \SI{50}{\hertz}, the inertia constant  $M_i = 2 W_{\mathrm{kin},i}/50$.} and $D_i\geq 0$ is the frequency dependent load at bus $i$, distributed according to \cref{tab:Nordic5_2}. In practice, $D_i$ is most likely unknown. Therefore, a conservative assumption is to assume that $D_i=0$ in the analysis.
The frequency-dependent actuator
\begin{equation}
    F_i(s) = H_i(s) K_i(s)
\end{equation}
represent a feedback controller $K_i(s)$ and a controllable actuator $H_i(s)$.
We consider two types of controllable actuators,  hydro and wind.

\begin{figure}[t!]
		\centering
		\includegraphics[width=\linewidth]{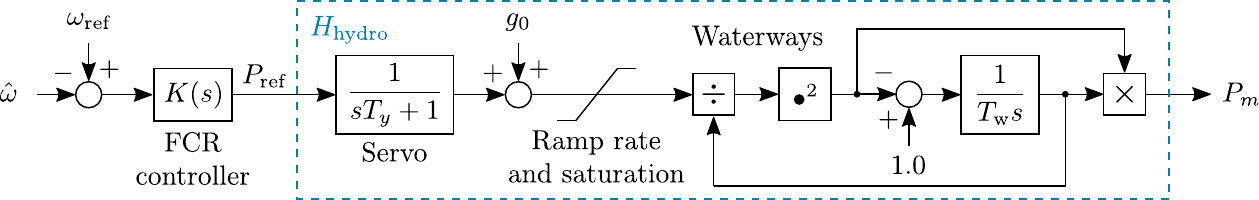}
		\caption{Block diagram of the hydro turbine and governor model.}
		\label{fig:hydro_block}
\end{figure}

The hydro governor implemented in this work is an adaptation of the model available in the Simulink Simscape Electrical library \cite{hydro-quebecSimscapeElectricalReference2020_manual}. It has been modified to allow for a general linear \ac{FCR} controller, $K(s)$, instead of the fixed PID/droop control structure, as shown in \cref{fig:hydro_block}. The servo rate limit is set to the default $\pm\SI{0.1}{\perunit\per\second}$. The nonlinear second-order model is useful for large-signal time-domain simulations.
For the linear analysis, the turbine is modeled as
\begin{equation}
\label{eq:Hhydro}
    H_\textrm{hydro}(s) = 
     2\frac{ z - s}{s + 2z}  
    \frac{1}{sT_y + 1} 
    , \quad z = \frac{1}{g_0 T_\mathrm{w}}
\end{equation}
where $T_y$ is the servo time constant, $g_0$ the initial gate opening, $T_\mathrm{w}$ the water time constant,  $\hat \omega$ the locally measured frequency, and $\omega_\mathrm{ref}$ the frequency reference. 

\begin{figure}[t!]
		\centering
		\includegraphics[width=\linewidth]{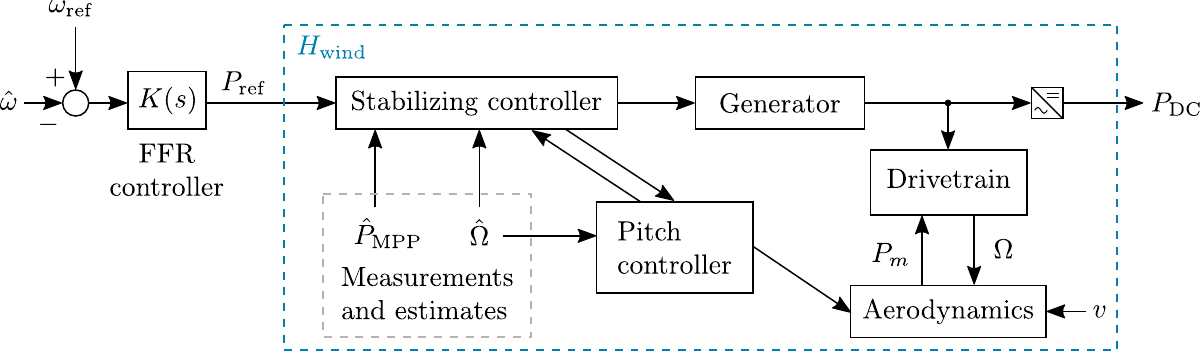}
		\caption{Block diagram of the wind turbine model.}
		\label{fig:wind_block}
\end{figure}

Wind turbines participating in \ac{FFR} are based on an adapted version of the \ac{NREL} \SI{5}{\mega\watt} baseline wind turbine model \cite{jonkmanDefinition5MWReference2009}. We consider uncurtailed operation below the rated wind speed. To allow for \ac{FFR} while tracking the \ac{MPP}, the control system has been modified according to \cite{bjorkVariablespeedWindTurbineunpublished} by adding a stabilizing feedback controller as illustrated in \cref{fig:wind_block}. For the linear analysis, the wind turbine is modeled as
\begin{equation}
    \label{eq:wind}
    H_\mathrm{wind}(s) = \frac{s - z}{s + k_\mathrm{stab} - z}, \quad z =  v \,  C_\Omega
\end{equation}
where $v$ is the wind speed, $k_\mathrm{stab}$ is the stabilizing feedback gain, and $C_\Omega$ is a variable that depends on how much the active power $P_m$ decreases when the rotor speed $\Omega$ deviates from the optimal rotor speed $\Omega_\mathrm{MPP}$. 
If we allow the turbine to operate down to \SI{80}{\percent} of $\Omega_\mathrm{MPP}$, then $C_\Omega \leq C_{0.8} = 5.8 \cdot 10^{-3}$. 
Setting the parameter $z = v \, C_{0.8}$ in \eqref{eq:wind} and $k_\mathrm{stab} = 2 v \,  C_{0.8}$ then gives us a linear representation that overestimates the negative phase shift of $H_\mathrm{wind}(j\omega)$. In this way, we can use the linear model for a conservative stability analysis. 

\begin{table}[t!]
	\centering
	\caption{Kinetic energy, load damping, and the network incidence parameters~\cref{eq:incidence parameters} for the \SI{110}{\giga\wattsecond} test case. 
	}
	\label{tab:Nordic5_2}
	\begin{tabular}{c|cccc}
		\hline
		Bus& 
		$W_\mathrm{kin}$ [\si{\giga\wattsecond}]& 
		$D$ [\si{\mega\watt\per\hertz}]&
		$\gamma/2\pi$ [\si{\giga\watt\per\radian}] \\
		\hline
		1 &34 &  150 & 6.2
		\\
		2 &  22.5 & 60 & 10.2
		\\
		3 & 7.5 &  20 & 5.2
		\\
		4 & 33 &  120 & 7.5
		\\
		5 &  13 &  50 & 3.0
		\\
		\hline
	\end{tabular}
\end{table}

\subsection{Hydro--FCR}
\label{sec:hydro_only_stab}
\begin{figure}[t!]
	\captionsetup[subfloat]{farskip=0pt}
	\centering
	\subfloat[\label{fig:poles_hydro_only}Open-loop poles of $\mathcal{G}'(s)$ and a modified Nyquist $\mathcal{D}_r$-contour that avoid any RHP poles slower than $r$.]{\includegraphics[scale=0.52]{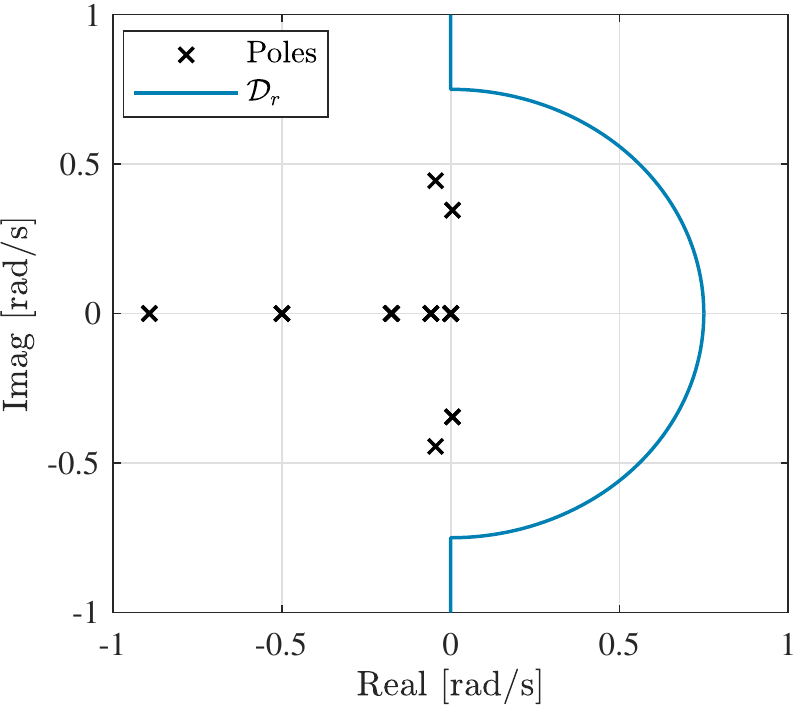}}	
	\hfill
	\subfloat[\label{fig:vertices_hydro_only}The image of vertices $\gamma_ig_i(s)\in\mathcal{G}'(s)$ as $s$ goes on the positive imaginary part of the $\mathcal{D}_r$-contour.]{\includegraphics[scale=0.52]{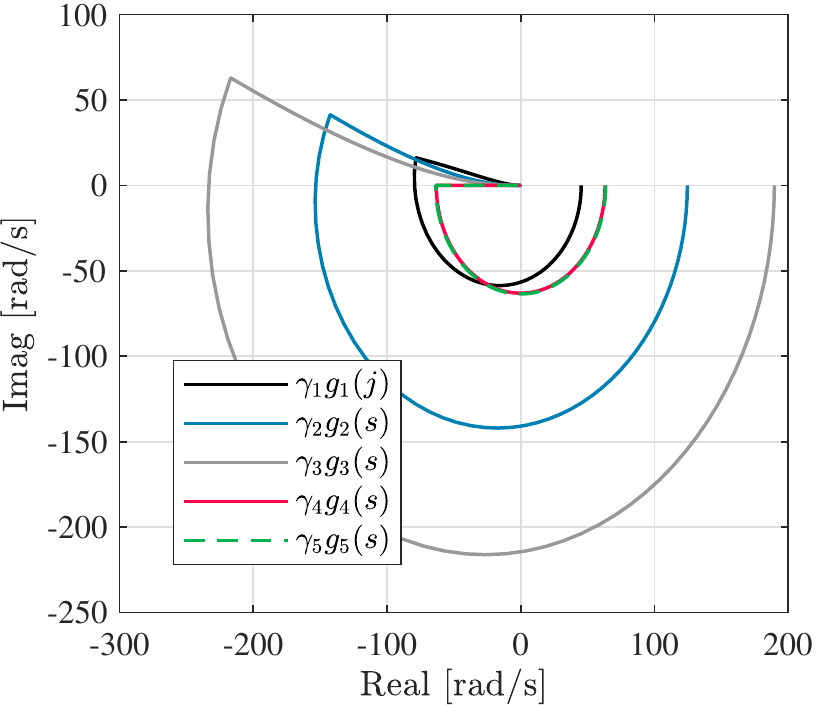}}	
	\caption{Open-loop poles and vertices of the N5 test system with hydro--\ac{FCR}, assuming that the frequency dependent load $D_i = 0$, $i\in\{1,\ldots,5\}$.}
	\label{fig:Nordic5_hydro_only1}
\end{figure}

Let us derive a decentralized stability criterion for a system in which controllable frequency reserves are solemnly provided by the hydro turbines at buses 1, 2, and 3. 
Let the parameter and \ac{FCR} resources be distributed according to \cref{tab:Nordic5_hydro} and 
let the \ac{FCR} feedback controller be tuned using the model matching method presented in \cite{bjorkDynamicVirtualPowerunpublished}. At bus $i$ we have
\begin{equation}
\label{eq:K_FCR_hydro}
    K_i(s) = c_i F_\mathrm{des}(s) \hat{H}_{\mathrm{hydro},i}^\inv(s)
\end{equation}
where the constant $c_i$ is the share of the total \ac{FCR} so that $c_1+c_2+c_3 = 1$. The transfer function $F_\mathrm{des}(s)$ is the \ac{FCR} design target \cref{eq:desiredKfcr}, and $\hat{H}_{\mathrm{hydro},i}(s)$ is a \ac{MP} estimate of \eqref{eq:Hhydro}. With \eqref{eq:K_FCR_hydro}, the controllable frequency-dependent actuator
\begin{equation}
    \label{eq:FCR_hydro}
    F_{\mathrm{hydro},i}(s) = K_i(s) {H}_{\mathrm{hydro},i}(s) = c_i F_\mathrm{des}(s) \frac{z_i - s}{z_i + s}.
\end{equation}
Since $H_\mathrm{hydro}(s)$ is \ac{NMP}, the agent
\begin{equation}
	\label{eq:agent_Nordic5}
	g_i(s) = \frac{1}{s} \frac{1}{s M_i+ F_{\mathrm{hydro},i}(s) + D_i}
\end{equation}
may have unstable poles. With the \ac{FCR} controller \eqref{eq:K_FCR_hydro}, the agents do in fact have unstable poles.
As shown in \cref{fig:poles_hydro_only} the unstable poles of agents \eqref{eq:agent_Nordic5} lies fairly close to the origin. Here, the unstable poles lie around \SI{0.50}{\radian\per\second}. The slowest interarea mode is known to be around \SI{0.37}{\hertz} ($\approx$ \SI{2.32}{\radian\per\second}). It is therefore quite easy to find a suitable modified Nyquist $\mathcal{D}_r$-contour that excludes the unstable poles. In \cref{fig:Nordic5_hydro_only1}, we choose $r=\SI{0.75}{\radian\per\second}$. To derive a decentralized stability criterion using \cref{cor:relax_syncronization_hetero}, we look at the field of values \cref{eq:field_value_cor}, spanned by vertices $\gamma_i g_i(s)$, as $s$ goes clockwise around the $\mathcal{D}_r$-contour. 

\begin{table}[t!]
\centering
\caption{Machine parameters for the \SI{110}{\giga\wattsecond} test case. 
Time constants and distribution of \ac{FCR} are based on the case study in \cite{saarinenFullscaleTestModelling2016}.
}
\label{tab:Nordic5_hydro}
\begin{tabular}{c|cccccc}
\hline
  Bus & $P_\mathrm{gen}$ [\si{\mega\watt}] & FCR [\si{\percent}] &
  $T_y$ & $T_\mathrm{w}$ & $g_0$ \\
 \hline
1 & \SI{9000} & 60 & 0.2 & 0.7 & 0.8 
\\
2  & \SI{6000} & 30 & 0.2 & 1.4 & 0.8 
\\
3  & \SI{2000} & 10 & 0.2 & 1.4 & 0.8 
\\
4 & \SI{5000} & -- & -- & -- & -- 
\\
5  & \SI{2000} & -- & -- & -- & -- 
\\
\hline
\end{tabular}
\end{table}

\paragraph{Load Damping Excluded}
Since we do not control the frequency dependent loads $D_i$, $i\in\{1,\ldots,5\}$, a reasonable conservative modeling assumption is to let $D_i=0$.
In \cref{fig:vertices_hydro_only} we see that we cannot derive a stability criterion with the proposed controller \eqref{eq:K_FCR_hydro} since the vertices that span the field of value approaches the origin from the top-left quadrant, thereby encircling  $-1$. In fact, the analysis suggests that the system is unstable, without other sources that contribute to damping. To amend this, we may either modify the \ac{FCR} controller \cref{eq:K_FCR_hydro} or assume that we know the frequency dependent loads $D_i>0$ and include these in the analysis. Here, we will use the latter, since this provides a good analogy to the case where we supplement hydro--FCR with wind--\ac{FFR} in \cref{sec:wind_and_hydro_stab}. 

\begin{figure}[t!]
	\captionsetup[subfloat]{farskip=0pt}
	\centering
	\subfloat[\label{fig:poles_hydro_and_loads}Open-loop poles of $\mathcal{G}'(s)$ and two modified Nyquist $\mathcal{D}_r$-contours with different radii.]{\includegraphics[scale=0.52]{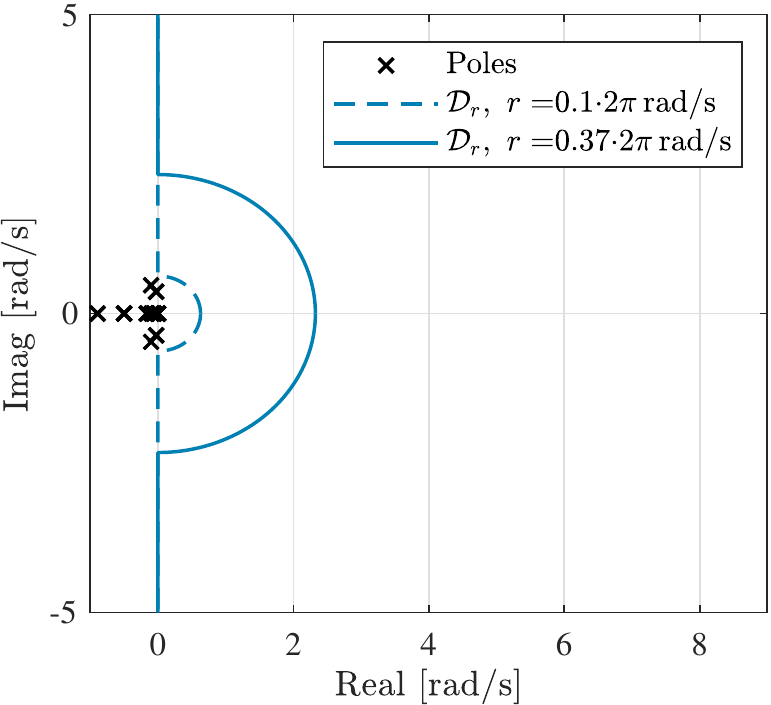}}	
	\hfill
	\subfloat[\label{fig:vertices_hydro_and_loads}The image of vertices $\gamma_ig_i(s)\in\mathcal{G}'(s)$ as $s$ goes on the positive imaginary part of the two $\mathcal{D}_r$-contours.]{\includegraphics[scale=0.52]{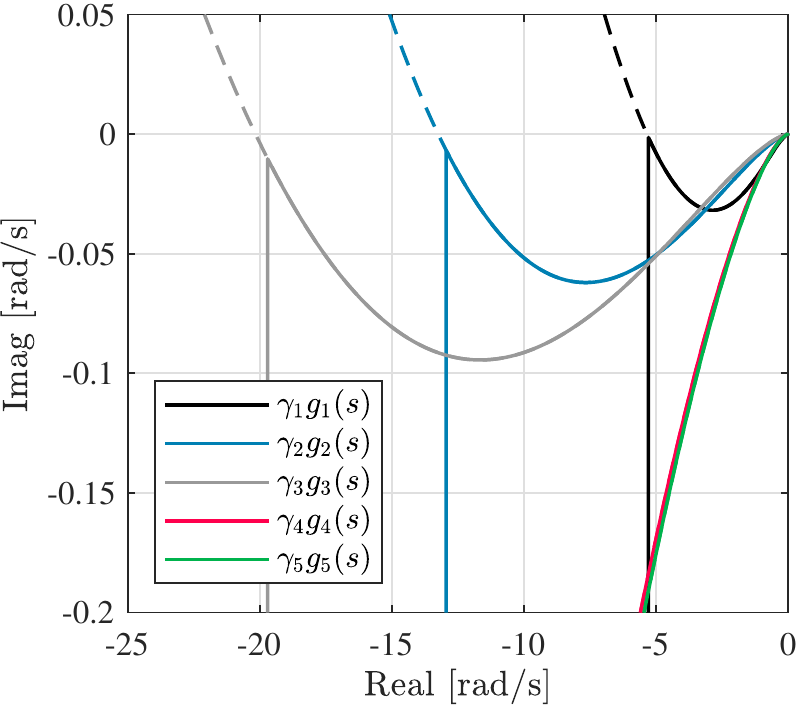}}	
	\caption{Open-loop poles and vertices of the N5 test system with hydro--\ac{FCR}, assuming that the distribution of the frequency dependent loads are known.}
	\label{fig:Nordic5_hydro_and_loads1}
\end{figure}

\paragraph{Load Damping Included}
Assume that the distribution of the frequency dependent loads shown in \cref{tab:Nordic5_2} is known and that we therefore can include these in the analysis. With $D_i>0$, the trajectories move towards the bottom left quadrant, at least for higher frequencies where $D_i>|F_{\mathrm{hydro},i}(j\omega)|$. As the vertices move towards the bottom left quadrant, they no longer encircle $-1$. However, we see that vertices 1, 2, and 3 still go back up into the top-left quadrant to the left of $-1$. Therefore, the system has a lower gain margin $\alpha > 0$. This is a problem since \cref{cor:syncronization_hetero,cor:relax_syncronization_hetero} requires that the field of values $\alpha \field\big(\mathcal{G}'(s)\big)$, $\alpha \in (0,1]$, does not encircle $-1$.
That is, none of the vertices are allowed to be left of $-1$ in the top left quadrant. Using \cref{cor:relax_syncronization_hetero} we can circumvent this problem by only looking on the image of vertices $\gamma_i g_i(s)$ for $s\in\mathcal{D}_r$. Setting the inner radius $r$ sufficiently large, we eventually find a point where the field of values cannot encircle $-1$. If frequency-dependent loads are distributed according to \cref{tab:Nordic5_2}, then we need $r$ to be larger than $0.37\cdot 2 \pi$\,\si{\radian\per\second}, as seen in \cref{fig:Nordic5_hydro_and_loads1}. \cref{cor:relax_syncronization_hetero} then tells us that we cannot have any unstable interarea modes with a eigenfrequency faster than \SI{0.37}{\hertz}, which we know to be the lower bound for the high-inertia operating condition. 
\begin{rem}
    For this low-inertia scenario, we know that the slowest interarea mode will have an eigenfrequency around  \SI{0.49}{\hertz}. We can therefore safely say that no interarea mode will be destabilized by the hydro--\ac{FCR}.
\end{rem}

\begin{figure*}[tb!]
\centering
\includegraphics[width=\textwidth]{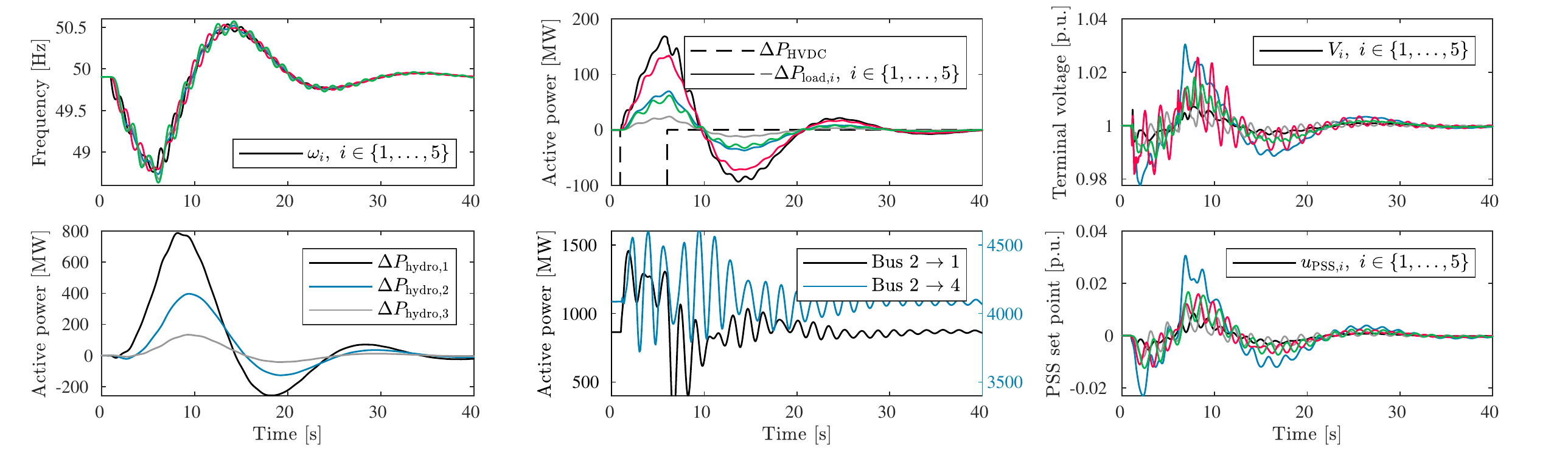}
\caption{Simulation of a \SI{5}{\second} disconnection of a \SI{1400}{\mega\watt} importing dc link in the N5 test system with hydro--\ac{FCR} \eqref{eq:FCR_hydro}.}
\label{fig_run_no_wind}
\end{figure*}

The simulated response to a \SI{5}{\second} disconnection of the \SI{1400}{\mega\watt} importing dc link is shown in \cref{fig_run_no_wind}. The frequency deviation (top left) is limited by the help of hydro--\ac{FCR} (bottom left) and the frequency-dependent loads (top center). The dc fault excites the north--south interarea mode, which is clearly visible on the tie-line flows (bottom-center). The stability of the system is aided by fast frequency-dependent loads, but network losses and voltage dynamics also play a role. For example, fast-acting excitation control, used to maintain the machine terminal voltage (top right), has a destabilizing effect on the interarea modes \cite{kundurPowerSystemStability1994}. To mitigate the destabilizing effect of the voltage control, \ac{PSSs} (bottom-right) have been installed on the machines.

\subsection{Hydro--{FCR} Supplemented with Wind--{FFR}}
\label{sec:wind_and_hydro_stab}
Let us now consider the case where we supplement hydro--\ac{FCR} with wind--\ac{FFR} at buses 1, 2, and 3. The obvious advantage of this approach is that the wind--\ac{FFR} is a conscious design choice. We do not have to base our stability analysis on assumptions about the load behaviour as we did in \cref{sec:hydro_only_stab}. Note that we have chosen a positive real design target \eqref{eq:desiredKfcr}, so 
\begin{equation}
   \Real\big( F_\mathrm{des}(j\omega)\big) \geq 0, \quad \forall \omega\geq 0.
\end{equation}
Using the model matching design proposed in \cite{bjorkDynamicVirtualPowerunpublished}, we could design a supplementary wind--\ac{FFR} so that the combined \ac{FCR} and \ac{FFR} at bus $i$ matches $c_i F_\mathrm{des}(s)$. As a result, all of the vertices would be negative imaginary
\begin{equation}
    \Imag \big(\gamma_i g_i(j\omega) \big) \leq 0, \quad \forall \omega\geq 0
\end{equation}
and therefore unable to encircle $-1$. In practice, however, it will likely not be possible to achieve strictly positive real frequency reserves. Even if we do not have dynamic limitations in the form of \ac{NMP} zeros, such as with hydro power, we will always have time delays. Here, we will show how \cref{cor:syncronization_hetero,cor:relax_syncronization_hetero} can be used to define a decentralized stability criterion even for a heterogeneous network with time delays.

To complement the hydro--\ac{FCR} we let the wind turbines at buses 1, 2, and 3 participate in \ac{FFR}. As mentioned earlier, this can be done using the model matching \cite{bjorkDynamicVirtualPowerunpublished}. To make the result comparable to \cref{sec:hydro_only_stab}, however, we design the wind--\ac{FFR} as a proportional frequency controller, making it comparable to the frequency-dependent load. Let 
\begin{equation}
\label{eq:FFR_wind}
    F_{\mathrm{wind},i}(s) = c_i k_\mathrm{FFR} \frac{5 s e^{-s\tau_i}}{5s +1} {H}_{\mathrm{wind},i}(s), 
\end{equation}
for $i \in\{1,2,3\}$, where we choose
$k_\mathrm{FFR} = \SI{1000}{\mega\watt\per\hertz}$. Since power outtake makes the wind turbine deviate from the \ac{MPP}, it cannot provide any sustained control action. This behaviour is captured by the all-pass characteristic in \eqref{eq:wind}. The turbines are able to achieve tight control at frequencies above \SI{0.06}{\radian\per\second}, i.e., they are well capable of damping interarea modes. Using a washout filter with a lower bandwidth of \SI{0.2}{\radian\per\second} we avoid steady control action. It is fairly straightforward to show that the wind farm and the hydro unit form a locally stable subsystem at bus 1, 2, and 3, respectively. For analyzing the global stability, the agents to consider then becomes
\begin{equation}
	\label{eq:agent_Nordic5_wind}
	g_i(s) = \frac{1}{s} \frac{1}{s M_i+ F_{\mathrm{hydro},i}(s) + F_{\mathrm{wind},i}(s)}
\end{equation}
for $i \in\{1,2,3\}$, and 
\begin{equation}
	\label{eq:agent_Nordic5_mach}
	 g_i(s) = \frac{1}{s} \frac{1}{s M_i}
\end{equation}
for $i \in\{4,5\}$. For simplicity, assume that the network incidence parameters $\gamma_i$, for the hydro--wind subsystems, are the same as in \cref{sec:wind_and_hydro_stab}. Let the wind speed and power rating be distributed according to \cref{tab:Nordic5_wind}. Furthermore, lets assume that the delay $\tau_i =\tau = \SI{100}{\milli\second}$ at all buses. Under these circumstances, neither of the vertices encircles the point $-1$, as seen in \cref{fig:Nordic5_hydro_and_wind1}. The proposed wind--\ac{FFR} is stronger than the frequency dependent load shown in \cref{fig:vertices_hydro_and_loads}. As a result, there is no risk for \emph{slow instability} due to vertices entering the top left quadrant to the left of $-1$.  Unlike the example with frequency-dependent loads, however, the vertices that correspond to agents with wind--\ac{FFR} cross over the real axis to the right of $-1$. The reason for this is the time delay in \eqref{eq:FFR_wind}. Consequently, we risk \emph{fast instability}. If we neglect the hydro--\ac{FFR}, then $\gamma_i g_i(s)$, $i \in\{1,2,3\}$ cross over the real axis exactly at $s = j \pi/2\tau$. This implies that a network with stronger connectivity (larger $\gamma_i$) will be more sensitive to fast instability caused by time delays.

\begin{table}[t!]
\centering
\caption{Wind turbine parameters for the \SI{110}{\giga\wattsecond} test case.
}
\label{tab:Nordic5_wind}
\begin{tabular}{c|cccc}
\hline
  Bus & $P_\mathrm{nom}$ [\si{\mega\watt}] & FFR [\si{\percent}] &
  $v$ [\si{\meter\per\second}] & $P_\mathrm{MPP}$ [\si{\mega\watt}]\\
 \hline
1 & \SI{1000} & 60 & 10 & 695 %& 1.19 
\\
2  & \SI{1000} & 30 & 6 & 150 %& 0.71 
\\
3  & \SI{500} & 10 & 7 & 120 %& 0.83 
\\
\hline
\end{tabular}
\end{table}

\begin{figure*}[tb!]
\centering
\includegraphics[width=\textwidth]{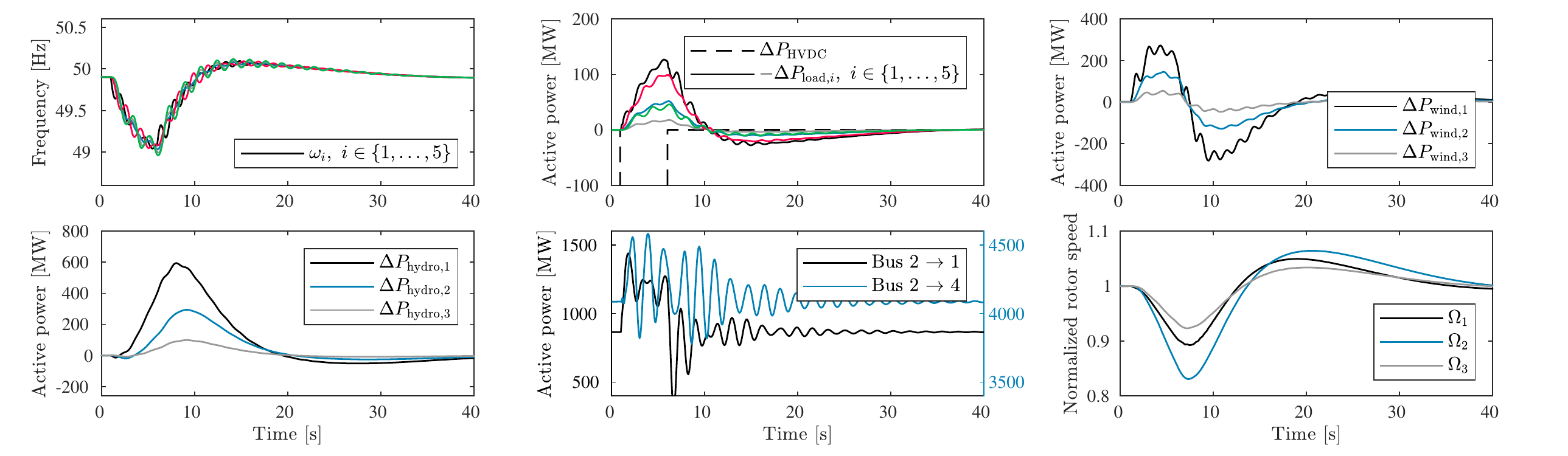}
\caption{Simulation of a \SI{5}{\second} disconnection of a \SI{1400}{\mega\watt} importing dc link in the N5 test system with hydro--\ac{FCR} \eqref{eq:FCR_hydro} and wind--\ac{FFR} \eqref{eq:FFR_wind}.}
\label{fig_run_with_wind}
\end{figure*}

The simulated response to a \SI{5}{\second} disconnection of the \SI{1400}{\mega\watt} importing dc link is shown in \cref{fig_run_with_wind}. Apart for the assisting wind--\ac{FFR}, the setup is identical to the system setup used in \cref{fig_run_no_wind}. As can be seen in the frequency response (top-left) and the north--south tie-line flows (bottom-center), the wind--\ac{FFR} (top-right) not only improves the frequency disturbance attenuation, but also improves the attenuation of interarea modes. The power excursion during \ac{FFR} decelerates the wind turbines (bottom-right). However, they are still within the allowed operating range, above \SI{80}{\percent} of the normalized rotor speed $\Omega/\Omega_\mathrm{MPP}$.

\subsection{Summary: A Scalable Nyquist Stability Criterion.}
\begin{figure}[t!]
	\captionsetup[subfloat]{farskip=0pt}
	\centering
	\subfloat[\label{fig:poles_hydro_and_wind}The image of vertices \cref{eq:agent_Nordic5_wind,eq:agent_Nordic5_mach} as $s$ goes on the positive imaginary part of the $\mathcal{D}_r$-contours in \cref{fig:poles_hydro_and_loads}.]{\includegraphics[scale=0.52]{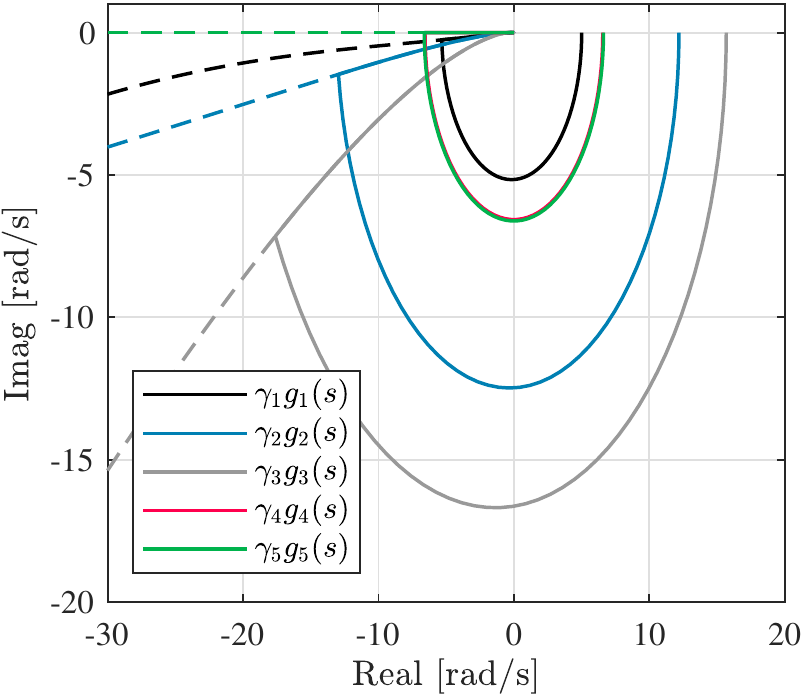}}	
	\hfill
	\subfloat[\label{fig:vertices_hydro_and_wind_zoom}Zoom in around $-1$. The ``$\times$" marks the value of vertex $\gamma_ig_i(j \omega)$ at $\omega =  \pi/2\tau$.]{\includegraphics[scale=0.52]{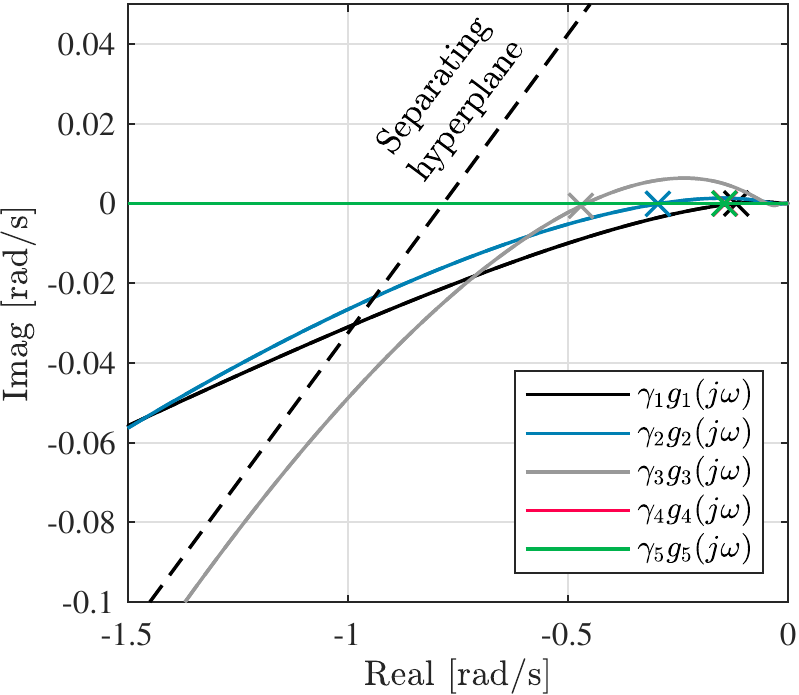}}	
	\caption{Vertices of the N5 test system with hydro--\ac{FCR} and wind--\ac{FFR}. }
	\label{fig:Nordic5_hydro_and_wind1}
\end{figure}

The following algorithm can be used as a blue print for a decentralized scalable Nyquist stability criterion.
\begin{itemize}
\item
Define a modified Nyquist $\mathcal{D}_r$-contour, a separating hyperplane\footnote{As an example, the dashed line in \cref{fig:vertices_hydro_and_wind_zoom} would be a suitable hyperplane for this network.}, and a time constant $\tau_\mathrm{max}$, to be used by all agents in the network.
\item We need an estimate of the lower bound for
the local inertia $M_i\geq 0$ and an estimate of
the controller and actuator dynamics $F_i(s)$ of devices that we want to connect.
We also need to know an upper bound on the local network incidence~parameter~$\gamma_i$.
 We then~require~that
\begin{enumerate}
    \item the agent $g_i(s)$ does not have unstable poles inside the modified Nyquist $\mathcal{D}_r$-contour,
    \item the vertex $\gamma_ig_i(s)$ does not enter the top-left quadrant to the left of $-1$ as $s$ goes along the positive imaginary part of the $\mathcal{D}_r$-contour, and that
    \item the vertex $\gamma_ig_i(j\omega)$ lies to the right of the hyperplane for all $\omega>\pi/2\tau_\mathrm{max}$.
\end{enumerate}
\end{itemize}
If every agent that connects to the network fulfills these criteria, then the network has no unstable interarea mode with an eigenfrequency faster~than~$r$. The frequency of slow system-wide interarea modes are easily observable from phasor measurements. The required bound for~$r$ should therefore be well known by the system operator. Note that in \cref{fig:vertices_hydro_and_wind_zoom}, agents $i\in\{4,5\}$ do not participate in frequency control. It is therefore impossible to base the stability criterion solemnly on a separating hyperplane in the Nyquist diagram. 

\section{Conclusions}
\label{sec:hetero:concl}
A scalable, decentralized stability criterion has been derived for a network with heterogeneous agents.
Stability is assessed without making prior assumptions on network losses or dynamics
by directly applying the generalized Nyquist criterion on the field of values spanned by the agents.
Using the proposed method, local stability criteria were derived for systems where passivity or separating hyperplane methods are impossible, e.g. if we have a mix of time-delayed actuators and uncontrolled agents. 
The results were validated in a detailed nonlinear power system model where we studied hydro--\ac{FCR} and wind--\ac{FFR}. It was shown that if actuators have slow \ac{RHP} zero dynamics (as is typically the case for hydro governors), then the local bus dynamics (the agent) may be unstable. As long as the frequency of the slowest interarea mode is known, it is possible to define a decentralized stability criterion, even for a network with unstable agents.

Typically, power system stability are separated into frequency, rotor angle, and voltage stability. These are typically treated separately. This work presents a unified framework for analyzing both frequency and rotor angle stability. The results show that we risk destabilizing the interarea modes if we demand bandwidth limited actuators, such as \ac{NMP} hydro turbines, to provide fast reserve power in low inertia power systems. It was shown that a convenient way to mitigate this problem is to allow converter-interfaced generation, capable of fast control action, to assist the conventional slow reserves.
For our future work, we will extend the result to second-order network dynamics so that we can include voltage dynamics in the stability analysis.

\bibliographystyle{IEEEtran}
\bibliography{refs_unpublished,refs_hetero}
\end{document}